\documentclass[12pt, a4paper, twoside, titlepage]{article}
\usepackage[latin1]{inputenc}
\usepackage[T1]{fontenc} 
\usepackage{graphicx}

\begin{document}
\title{Robust multiple comparisons against a control group with application in toxicology}
\date{May 2019}
\author{Ludwig A. Hothorn\\
Im Grund 12, D-31867 Lauenau, Germany and \\ Felix M. Kluxen\\ADAMA Deutschland GmbH, D-51149 Cologne, Germany}

\maketitle

\begin{abstract}
The Dunnett procedure compares several treatment or dose groups with a control group, while controlling the familywise error rate.
When deviations from the normal distribution and heterogeneous variances occur, the nominal $\alpha$ level may be violated, and power may be reduced.
Various robust modifications are discussed, whereby the novel most likely transformation (MLT)-Dunnett version is recommended as almost always appropriate by means of a simulation study. The MLT-Dunnett is especially useful because it can jointly and comparably analyse differently scaled endpoints. Furthermore, a related multiple endpoints test is proposed using the odds ratio as a common effect size. With the statistical software R, the method is readily applicable using the CRAN libraries \verb|multcomp| and \verb|mlt|, real data can be easily analyzed.

\end{abstract}

\section{Introduction}
The Dunnett \cite{Dunnett1955} multiple comparison procedure  is commonly used for treatment or dose groups inference against a placebo group in randomized clinical trials or against negative control in non-clinical studies.
This test belongs to a class of max-t multiple contrast tests for a one way, k-sample design assuming a common variance estimator $S$, and a common degree of freedom $df$, i.e. $N(\mu_i, \sigma^2)$ \cite{Hothorn2007}. Modifications in the generalized linear model are available using a general parametric model \cite{Hothorn2008}, particularly for proportions \cite{Schaarschmidt2009a}, poly-k estimates in long-term carcinogenicity studies \cite{Schaarschmidt2008}, the Cox-model \cite{Herberich2012} and multinomial vectors \cite{Schaarschmidt2017}.\\

Multiple deviations from the normal distribution occur frequently and simultaneously in real data; for example a continuous primary endpoint with common mildly unbalanced designs, such as  radiographical change in  joint space width when treating knee osteoarthritis \cite{Reginster2013} or hemoglobin change from baseline when characterizing anemia in chronic kidney disease \cite{Martin2017}.

Pre-tests on these conditions, such as Kolmogorov-Smirnov test, are not recommended \cite{Hothorn2019} and therefore robust modifications are considered in the current manuscript which are less sensitive to violations of the $N(\mu_i, \sigma^2)$ assumption, i.e. maintain $\hat{\alpha}$ and show no substantial power loss. \\
 In addition to robustness, there is usually the problem to encounter multiple differently scaled endpoints within one bioassay, such as clinical chemistry endpoints in a 13-week toxicological bioassay with sodium dichromate dihydrate administered to 
F344 rats \cite{NTP2002}. It is unrealistic to assume that all these endpoints follow a certain distribution (like the normal distribution) or can be converted to it by exactly one transformation (like commonly-used log transformation). Therefore, a test is needed which evaluates differently scaled endpoints fairly comparably using either an independent univariate analysis or a simultaneous multiple endpoint analysis.\\
The availability of simultaneous confidence intervals (in addition to compatibly adjusted p-values) should be ensured for any testing procedure, because estimation procedures are more informative than statistical tests which lead to a practical limitation of single-step procedures. For example, the ICH E9 guideline  states that \textit{'Estimates of treatment effects should be accompanied by confidence intervals...'}.

Five modifications of the Dunnett test will be compared by the evaluation of a real data example and a simulation study with the original test, and to each other, particularly considering small sample size design (common in non-clinical studies, down to $n_i=10$): i) nonparametric procedure for relative effect size \cite{Konietschke2012}, ii) Dunnett-test using modified degree of freedom according to Sattherwhaite \cite{Hasler2008}, iii) Dunnett-test using sandwich estimator for robust variance-covariance estimation \cite{Herberich2010}, iv) robust M-estimator \cite{Koller2011}, and v) a novel most likely transformation (MLT) \cite{Hothorn2018}. \\
Of those, the MLT-Dunnett test procedure performs best and can be recommended as being appropriate in conditions that are considered unsuitable for the original Dunnett test and is more powerful than the published alternatives.\\
Further, odds ratios (OR) can be used as a common effect size for a Dunnett-type procedure for both univariate endpoints and correlated multiple and differently scaled endpoints, optimally dichotomized for continuous endpoints by continuous outcome logistic regression (COLR, \cite{Lohse2017}). The R-code to implement these procedures is given alongside.

\section{Robust Dunnett-type procedures}
The multiple comparison procedure introduced by Dunnett \cite{Dunnett1955} estimates  a studentized tests statistic assuming normal distributed errors with homogeneous variances. Adjusted p-values and/or compatible simultaneous confidence limits are available. 
The simultaneous confidence intervals (sCI) (2-sided) of multiple contrast tests are
$[\sum_{i=0}^k c_i\bar{x}_i \pm S* t_{q,df,\mathbf{R},2-sided,1-\alpha}\sqrt{\sum_i^k c_i^2/n_i}]$ with the contrast coefficients $c_i$ (-1 for the control, 1 for a treated group, 0 otherwise for the Dunnett-test), $q$ the number of multiple contrasts (where $q=k$ for Dunnett-test), the correlation matrix $\mathbf{R}=(\rho_{ij})$ (for the Dunnett-test  $\rho_{ij}=\sqrt{\frac{1}{(1+n_{0}/n_{i})(1+n_{0}/n_{j})}} \quad (1 \leq i\neq j\leq k)$)(may be complex in other scenarios \cite{Hothorn2015}).  By means of the CRAN R package \verb|multcomp| \cite{Hothorn2008} these sCIs can be easily estimated, as well as the compatible, multiplicity-adjusted \textit{p}-values derived as an alternative.  In the case of variance heterogeneity (particularly in unbalanced designs), the common degree of freedom $df=\sum_{i=0}^{k}(n_{i}-1)$ can be replaced by $ df^{Sattherwhaite}$ and $\rho_{ij}$ depends on the group-specific variances estimates \cite{Hasler2008}. Alternatively, the common variance-covariance can be estimated via sandwich estimator \cite{Herberich2010}.\\
Although pairwise-ranking, such as Steel procedure, and joint-ranking, such as Kruskal-Wallis test are commonly used, a robust approach against violations of normal distribution and variance homogeneity should be used for the relative treatment effect between control (0) and treatment $i$: 
$p_{0i}=Pr(X_0 < X_i)+\frac{1}{2} Pr(X_0 = X_i)= \int{F_0dF_i}$ \cite{Konietschke2012}. 
The simultaneous confidence intervals are similar to the parametric standard above, where the correlation matrix is estimated from sample sizes, contrast coefficients and the sample specific variances. An appropriate t-distributed version and a range $[0,1]$ preventing transformations can be recommended. A related CRAN R package \verb|nparcomp| \cite{Konietschke2015} is available.
A robust procedure can be achieved by using robust estimators, such as the M estimator \cite{Koller2011}.  
The analysis of quite differently-scaled endpoints (including tied, skewed or censored) can be performed by the concept of most likely transformation \cite{Hothorn2018}.

The most likely transformation approach is embedded in a maximum likelihood framework whereas the parametrization of the monotone increasing transformation function is achieved by
Bernstein polynomials (of order 5 in the example). For a deeper discussion of this MLT- approach, we
refer the reader to \cite{Hothorn2018}.
The object-oriented properties of packages \verb|mlt| and \verb|multcomp| make it easy to estimate corresponding simultaneous Dunnett-type confidence intervals on the assumption of large $n_i$. The R code of all procedures is available below. \\

\paragraph{Real data example} Eleven clinical chemistry endpoints from a 13-week study with sodium dichromate dihydrate administered to F344 rats are used as an example \cite{NTP2002}. In Table \ref{tab:p250} multiplicity adjusted p-values for the comparison of the medium 250 mg dose versus control of the 11 endpoints are shown, i.e. original Dunnett (Dun), M-robust version (Rob), MLT-modification (MLT), non-parametric version (Rel). And, a normal/non-normal decision based on a Cramer test is given along with each comparison. The endpoint creatin kinase (CreatK) is interesting: for this non-normal distributed endpoint, the parametric Dunnett-test is non-significant, whereas the robust and the MLT modification reveal much smaller p-values. The response against dose groups in mg/kg bw/d is plotted as a boxplot in Figure \ref{fig:CreatK} and includes superimposed individual values next to mean and standard deviation. Note the extreme values in group 62.5 and 1000 mg/kg bw/d, which induce variance heterogeneity.  Although the comparison of interest 250 vs control reveals no extreme value and no obvious variance heterogeneity, it is not surprising that the Dunnett test reveals a large non-significant p-value because it is a k-sample test with a common mean square error estimator.

\begin{figure}[h]
	\centering
		\includegraphics[width=0.62\textwidth]{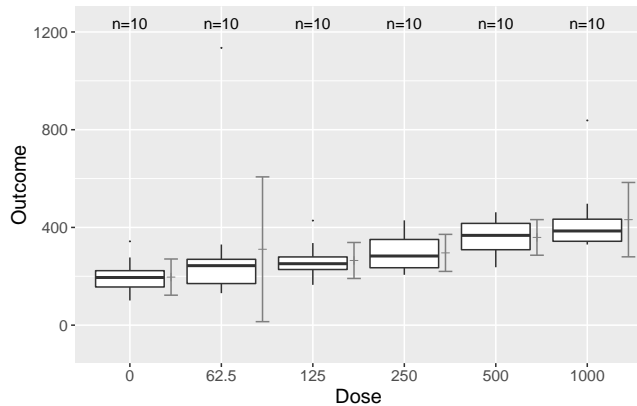}
	\caption{Creatin kinase example}
	\label{fig:CREA}
\end{figure}

\begin{table}[ht]
\scriptsize
\centering
\caption{Univariate 250 vs. control comparisons: adjusted p-values}
	\label{tab:p250}
\begin{tabular}{rrrrr|r}
  \hline
 & Dun & Rob & MLT & Rel & Normal? \\ 
  \hline
Nuc &  0.404 & 0.368 & 0.240 & 0.261 &yes\\ 
  BUN & 0.040 & 0.051 & 0.011 & 0.018 &yes \\ 
  SerG & 0.489 & 0.410 & 0.330 & 0.466 &no \\ 
  CreatK & 0.437 & 0.012 & 0.010 & 0.027 &no \\ 
  ALT & 0.005 & 0.001 & 0.000 & 0.000 & no \\ 
  SDH & 0.015 & 0.006 & 0.000 & 0.000 & no \\ 
  Chol & 0.077 & 0.079 & 0.026 & 0.079 &yes \\ 
  Tri &  0.001 & 0.003 & 0.001 & 0.029 & no \\ 
  Chlor & 0.834 & 0.679 & 0.740 & 0.740 &no \\ 
  Sodium & 0.97 & 0.99 & 0.97 & 0.99 &yes \\ 
  Potsa  & 0.71 & 0.99 & 0.91 & 0.99 & no \\ 
   \hline
	\label{tab:d2}
\end{tabular}
\end{table}

The R-code for the five modified Dunnett tests for the endpoint CreatK is:
\scriptsize
\begin{verbatim}
#### R code for the data example
library("devtools")
install_github("lahothorn/SiTuR")
data("clin", package="SiTuR") # import data example
clin$dose<-as.factor(clin$Dose) # dose as factor

library("mlt"); library("nparcomp")
library("robustbase"); library("ggplot2")
modCK<-lm(CreatKinase~dose, data=clin) # linear model
rmodCK<-lmrob(CreatKinase~dose, data=clin, 
              setting = "KS2014") # robust M estimators
CDU<-fortify(summary(glht(modCK, 
     linfct = mcp(dose = "Dunnett"))))# Dunnett test
rCDU<-fortify(summary(glht(rmodCK, 
      linfct = mcp(dose = "Dunnett")))) # robust Dunnett
yvar <- numeric_var("CreatKinase", support = 
        quantile(clin$CreatKinase, prob = c(.01, .99))) # MLT
bstorder<-5 # order of Bernstein polynomial
yb <- Bernstein_basis(yvar, ui = "increasing", 
                order =bstorder) # Bernstein polynomial
ma <- ctm(yb, shifting = ~ dose, 
      todistr = "Normal", data = clin) # condit transf mod
m_mlt<-mlt(ma, data = clin) # most likely transformation
K <- diag(length(coef(m_mlt))) # contrast matrix
rownames(K) <- names(coef(m_mlt))
matr<-bstorder+1
K <- K[-(1:matr),] # for order 5 Bernstein
C<-glht(m_mlt, linfct = K) # MLT-Dunnett-type test
CMLT<-fortify(summary(C)) 
NPC<-nparcomp(CreatKinase~dose, data=clin,
     asy.method = "probit", alternative = "two.sided", 
		 type = "Dunnett",plot.simci = FALSE, 
    info = FALSE)$Analysis$p.Value # relat effects 
pCK<-cbind(CDU[, c(1,6)], rCDU[, 6], CMLT[, 6], NPC)
\end{verbatim}
\normalsize

The object pCK contains the multiplicity-adjusted p-values for the modified Dunnett-tests. Notice, instead of the asymptotic version, for the common small sample sizes designs, a t-distributed maximum test can be recommended, simply by using the degree of freedom of the linear model\\
\scriptsize \verb|sC<-glht(m_mlt, linfct = K, df=modCK$df.residual).|
\normalsize For both randomized clinical trials and toxicological bioassays a directional evaluation, i.e. one-sided confidence limits (or compatible adjusted p-values) are appropriate and available.

\paragraph{A continuous outcome logistic regression approach} In addition to robustness, there is usually the problem to encounter multiple differently scaled endpoints within one bioassay: normally distributed (such as body weight), skewed distributed (such as ASAT), dichotomous (such as tumor rate), ordered categorized (such as graded pathological findings), etc. Usually different test principles are used, such as parametric, nonparametric and those for proportions. An alternative is the comparable evaluation of differently scaled endpoints with an approach that uses an uniform effect size. Continuous outcome logistic regression (COLR) \cite{Lohse2017} provides the dimensionless odd ratios (and their confidence intervals) as effect size optimally dichotomized for the endpoint specific distribution (any continuous, any discrete up to binary, censored etc.) as well as taking also the higher moments (location, scale and shape) into account. This approach is similar to a specific version of \textit{most likely transformation} (MLT) with the distribution function argument \textit{logistic} \cite{Hothorn2018} whereas the CRAN package \verb|tram| is used here. Against a recommendation for a routine use speaks its problematic small sample size characteristics of this asymptotic procedure.
A Dunnett-type test for the non-normal distributed endpoint creatine kinase is shown as an example:

\footnotesize
\begin{verbatim}
data("clin", package="SiTuR") # import data example
clin$dose<-as.factor(clin$Dose) # defining dose as factor
library("tram")
COC<-Colr(CreatKinase~dose, data=clin) # COLR
CC<-glht(COC, linfct = diag(length(coef(COC))))
ccMLT<-1/exp(confint(CC)$confint)# OR and sCI
summary(CC)$test$pvalue # adj.p-value
\end{verbatim}
\normalsize

The odds ratios versus control together with their lower simultaneous confidence limits and the adjusted p-values are given in Table 
\ref{tab:pp}:
 
\begin{table}[]
\small
\caption{Dunnett-type continuous outcome logistic regression}
\label{tab:pp}
\begin{tabular}{l|ll|l}
 Comparison& OR  & lower limit  & adj. p-value  \\ \hline
62.5/0 & 3 & 0.4 &  0.5  \\
125/0 & 6 &   0.8 & 0.11  \\
250/0 & 12 &  1.4 & 0.016\\
500/0 & 52 &  5.3 & 0.0001\\
1000/0 &  98&   9.7& 0.000002 \\
\end{tabular}
\end{table}

The chance of increased creatine kinase levels in the 250 mg/kg dose group is at least 1.4 times higher compared to control, which is an effect size representation equivalent to the p-value of 0.016.

\paragraph{Joint analysis of multiple correlated and differently-scaled endpoints}
To limit false positive claims it can be helpful to conduct a joint analysis of multiple endpoints considering their correlations and dimension $k$. 	 A related Dunnett-type test assuming multivariate normality and variance homogeneity \cite{Hasler2011} and variance heterogeneity \cite{Hasler2018} is available. A multiple endpoint approach based on the unique effect size odds ratio for differently scaled endpoints is proposed here. The joint distribution of the underlying Tmax-test \cite{Hasler2011} is achieved by multiple marginal model approach  \cite{Pipper2012}, which allows maximum tests on multiple linear models without the explicit formulation of the correlation matrix. The variance-covariance matrix of parameter estimates can be obtained using derivatives of the log-likelihood function in a single endpoint model, and hence for multiple models a vector of log-likelihood functions is used.  By plugging in parameter estimates, a consistent sandwich estimator of the variance-covariance matrix is obtained, i.e. the empirical covariance based on functions of the data, not on the data itself. A related function \verb|mmm| is available with the R package \verb|multcomp| \cite{Hothorn2008}. From the above clinical chemistry example, the bivariate case of creatine kinase and the right-skewed distributed alanine aminotransferase (ALT) is selected:

\scriptsize
\begin{verbatim}
yCK <- numeric_var("CreatKinase", support = 
       quantile(clin$CreatKinase, prob = c(.01, .99)))
yAL <- numeric_var("ALT", support = 
       quantile(clin$ALT, prob = c(.01, .99)))
bstorder<-5 # order of Bernstein polynomial
yC <- Bernstein_basis(yCK, ui = "increasing", 
      order =bstorder) # Bernstein polynomial
yA <- Bernstein_basis(yAL, ui = "increasing", 
      order =bstorder) # Bernstein polynomial
mC <- ctm(yC, shifting = ~ dose,todistr = "Logistic", 
      data = clin) # condit. transf. model
mC_mlt<-mlt(mC, data = clin) # most likely transformation
K <- diag(length(coef(mC_mlt))) # contrast matrix
rownames(K) <- names(coef(mC_mlt))
matr<-bstorder+1
K <- K[-(1:matr),] # matrix for order 5 Bernstein
C<-glht(mC_mlt, linfct = K) # MLT-Dunnett-type test
CMLT<-fortify(summary(C)) 

mA <- ctm(yA, shifting = ~ dose,todistr = "Logistic", 
      data = clin) # condit transf mod
mA_mlt<-mlt(mA, data = clin) # most likely transformation
A<-glht(mA_mlt, linfct = K) # MLT-Dunnett-type test
AMLT<-fortify(summary(A)) 

MMMDF <- c(mC_mlt$df, mA_mlt$df)
conLH <- diag(2)%x%K # contrast matrix for bivar Dunnett
bread.mlt_fit <- function(x) vcov(x) * nrow(x$data) # sandwich 
MMn <-summary(glht(mmm(mC_mlt, mA_mlt), 
			linfct=conLH, df=mean(MMMDF)))
\end{verbatim}
\normalsize
 
The related adjusted p-values (adjusted against multiple comparisons versus control and the two correlated endpoints) are are presented in Table \ref{tab:exa1}:

\begin{table}[ht]
\centering
\caption{Adjusted p-values bivariate MLT-Dunnett} 
\label{tab:exa1}
\begingroup\small
\begin{tabular}{lrr}
  \hline
Hypothesis & Lower limit OR & $p$-value \\ 
  \hline
Creatin: 62.5/0 & 0.2 & 0.6922 \\ 
 Creatin: 125/0 & 0.5 & 0.2354 \\ 
 Creatin: 250/0  & 0.8 & 0.0699 \\ 
 Creatin: 500/0   & 3.0 & 0.0069 \\ 
 Creatin: 1000/0  & 5.8 & 0.0023 \\ 
 ALT: 62.5/0  & 91.5 & 0.0004 \\ 
  ALT: 125/0  & 9.7 & 0.0028 \\ 
  ALT: 250/0  & 15.0 & 0.0016 \\ 
   ALT: 500/0 & 10.9 & 0.0023 \\ 
 ALT: 1000/0  & 18.2 & 0.0013 \\ 
   \hline
\end{tabular}
\endgroup
\end{table}
As expected, the multiplicity-adjusted p-values of the  bivariate analysis are increased compared to the univariate analysis (just as the lower limit is decreased analogously). This is the necessary penalty for a claim accross multiple endpoints, which increases with the number of considered endpoints and with lower correlations. Thus, a careful selection of relevant endpoints is suggested for a follow-up multivariate analysis.

\section{Simulations}
In a simulation study, the empirical size $\widehat{\alpha}$ under the null hypothesis ($H_0$) and the empirical power under the alternative hypothesis ($H_1$) of the following Dunnett-type procedures are compared (in 10000 runs): i) original 
\cite{Dunnett1955} (Dun), ii) adjusted for variance heterogeneity by sandwich estimator \cite{Herberich2012} (SaW), iii) adjusted for variance heterogeneity by reduced df (Sat), \cite{Hasler2008}, iv) robust linear model \cite{Koller2011} (Rob), v) most likely transformation \cite{Hothorn2018} (MLT), vi) pairwise-ranking Steel test \cite{Steel1959} (Ste), vii) nonparametric multiple contrast test \cite{Konietschke2015} (Rel). The expected values and variances are chosen similar to the endpoint cholesterol in the example study whereas a balanced and unbalanced design with a control and three dose groups with small sample sizes ($n_i=10)$ were considered for normal distribution (N) and a skewed distribution (M), which consist of a mixture of $10\%$  and $20\%$ extreme high values respectively, parametrized by variance inflation factor of $\xi$.


The results of the simulation study are shown in Table \ref{tab:sim2}. Under the null-hypothesis the empirical $\alpha$ levels should be near the the nominal level of $0.05$: too large values (liberal behavior) are a sign against the validity of a level $\alpha$-test, too small values indicates a conservative behavior, which leads to undesirable high false negative rate in safety assessment. The power of a test must and cannot be maximal in every situation examined, but it should be uniformly high. 

\begin{table}[ht]
\tiny
\caption{Simulation results: $\widehat{\alpha}$ and power}
\label{tab:sim2}
\centering
\begin{tabular}{rrrr|rrrrrrrr}
  \hline
Hyp & $\xi$ & $n_1$ & $n_4$ & Dun & Sat & SaW & Rob & MLT & Ste & Rel \\ 
  \hline
 N,H0 & 1 & 10 & 10 & 0.05 & 0.03 & 0.08&   0.06 & 0.06 & 0.04 & 0.03  \\ 
H0 & 2 & 10 & 10 & 0.04 & 0.03 & 0.08 &  0.06 & 0.07 & 0.04 & 0.03 \\ 
H0 & 3 & 10 & 10 & 0.04 & 0.02 & 0.07 &  0.05 & 0.05 & 0.04 & 0.04  \\ 
H0 & 4 & 10 & 10 & 0.03 & 0.03 & 0.06 &  0.04 & 0.06 & 0.03 & 0.03  \\ 
\hline
H1 & 1 & 10 & 10 & 0.84 & 0.79 & 0.89 &  0.86 & 0.85 & 0.70 & 0.74  \\ 
H1 & 2 & 10 & 10 & 0.75 & 0.69 & 0.81 &  0.79 & 0.77 & 0.63 & 0.65  \\ 
H1 & 3 & 10 & 10 & 0.65 & 0.58 & 0.69 &  0.72 & 0.70 & 0.57 & 0.56  \\ 
H1 & 4 & 10 & 10 & 0.52 & 0.47 & 0.58 &  0.71 & 0.63 & 0.51 & 0.47  \\ 
\hline
H0 & 1 & 20 & 10 & 0.04 & 0.03 & 0.07 &  0.06 & 0.05 & 0.04 & 0.04  \\ 
H0 & 4 & 20 & 10 & 0.06 & 0.03 & 0.06 &  0.06 & 0.06 & 0.04 & 0.04  \\ 
\hline
H1 & 1 & 20 & 10 & 0.94 & 0.92 & 0.94 &  0.94 & 0.95 & 0.90 & 0.88  \\ 
H1 & 4 & 20 & 10 & 0.73 & 0.67 & 0.65 &  0.84 & 0.78 & 0.72 & 0.62  \\ 
   \hline	
H0 & 1 & 5 & 20 & 0.04 & 0.02 & 0.11 &  0.05 & 0.05 & 0.03 & 0.05 \\
H0 & 4 & 5 & 20 & 0.02 & 0.00 & 0.11 &  0.05 & 0.04 & 0.03 & 0.06  \\ 
\hline
H1 & 1 & 5 & 20 & 0.64 & 0.47 & 0.79 &  0.67 & 0.70 & 0.50 & 0.63  \\ 
H1 & 4 & 5 & 20 & 0.36 & 0.22 & 0.67 &  0.63 & 0.57 & 0.41 & 0.56  \\
\hline
\hline

M,H0 & 1 & 20 & 10 & 0.09 & 0.07 & 0.09&  0.16 & 0.08 & 0.06 & 0.06 \\ 
H0 & 2 & 20 & 10 & 0.09 & 0.07 & 0.08 &  0.20 & 0.07 & 0.04 & 0.05 \\ 
H0 & 3 & 20 & 10 & 0.13 & 0.10 & 0.09 &  0.22 & 0.09 & 0.07 & 0.05 \\ 
H0 & 4 & 20 & 10 & 0.13 & 0.09 & 0.07 &  0.20 & 0.07 & 0.06 & 0.05 \\ 
\hline
H1 & 1 & 20 & 10 & 0.71 & 0.68 & 0.57 &  0.74 & 0.65 & 0.51 & 0.34 \\ 
H1 & 2 & 20 & 10 & 0.63 & 0.57 & 0.44 &  0.69 & 0.53 & 0.43 & 0.26 \\ 
H1 & 3 & 20 & 10 & 0.52 & 0.47 & 0.33 &  0.64 & 0.44 & 0.37 & 0.22 \\ 
H1 & 4 & 20 & 10 & 0.46 & 0.40 & 0.26 &  0.61 & 0.39 & 0.34 & 0.19 \\ 
\hline
H0 & 1 & 5 & 20 & 0.00 & 0.00 & 0.10 &  0.02 & 0.01 & 0.00 & 0.04  \\ 
H0 & 4 & 5 & 20 & 0.00 & 0.00 & 0.09 &  0.02 & 0.01 & 0.00 & 0.05  \\ 
\hline
H1 & 1 & 5 & 20 & 0.31 & 0.19 & 0.59 & 0.41 & 0.32 & 0.21 & 0.41  \\ 
H1 & 4 & 5 & 20 & 0.10 & 0.05 & 0.41 &  0.35 & 0.20 & 0.15 & 0.34  \\ 
\hline
\end{tabular}
\end{table}
The simulations show that the problematic conservative behavior with increasing variance heterogeneity only occurs when $n_i<n_0$.
As expected, the asymptotic tests (SaW, MLT) reveal a liberal behavior for the small $n_i$, common in toxicology.
Increasing variance in the high dose group causes a problematic power loss, which is still lowest at MLT. For rarely used designs with $n_0 <n_i$ Dun and Sat are extremely conservative (connected with power loss). As expected for mixing distribution, a substantial power loss occurs which is extreme for Rel. Even from a size/power perspective,  MLT is an almost always appropriate test.


\section{Discussion}
Multiple deviations from the normal distribution occur frequently and simultaneously in real data. This is often caused by extreme single values and heterogeneous variances and may be complicated by unbalanced designs.  In these cases, the Dunnett procedure, which is recommended in many guidelines, can react with size violation, i.e. deviation from the nominal alpha, and/or power loss. 

There are two strategies to address this. Common is an application of so-called assumption/pre- tests, which are subsequently used in a decision tree-like procedure. There are several issues with this approach, see for details \cite{Hothorn2019}. One major disadvantage is the use of different test types in the final main analysis, for which effect sizes and estimates are not present or comparable. Another problem is that the main test alternatives do not necessarily address the issues identified by the pre-tests, e.g. heterogeneous variances.  The method presented in this manuscript allows the application of a robust test, i.e. a common and comparable analysis for multiple endpoints, which is applicable even when deviations from the Dunnett test assumptions occur.   Various robust modifications have been compared, whereby the MLT-version is recommended as almost always appropriate. As shown in a simulation study, all procedures considered are problematic for very small $n_i$, especially for unbalanced designs (e.g. the common $n_0>n_i$ ) and variance heterogeneity- both due to insufficient power and non-maintenance of the intended $\alpha$ level. 
While the MLT-Dunnett performed best in conditions of the simulation study, no suitable test was identified for common small sample designs, and accordingly statistical significance needs to be considered cautiously in the absence of biological relevance. 

We showed the application of the MLT-Dunnett in real data examples, namely the derivation of p-values and effect sizes. Its performance has been compared in a simulation study with the original Dunnett test and its various published alternatives and modifications.

The modification of using continuous outcome logistic regression (COLR) allows the comparison of common odds ratios over differently scaled endpoints, which is a clear advantage to using different main tests for those endpoints. Since such endpoints usually occur together in toxicological bioassays, this procedure might be helpful to assess a toxic response in more detail and coherence. The application of multiple marginal models allows simultaneous claims over several correlated endpoints, which allows a more refined statistical, and by the use of odds ratios, toxicological assessment of correlated effects. The current practice in toxicology is to statistically compare each endpoint within one bioassay in isolation and thus consciously ignoring known correlations while accepting size violation along with increased rate of false positive claims.

Together, the MLT-Dunnett can be recommended to be used in toxicology especially because several, differently scaled endpoints can be analyzed comparably and jointly as a multiple endpoint test. Both adjusted p-values and simultaneous confidence intervals are available for all procedures, although some of the latter can no longer be interpreted on the clinical scale. With the CRAN libraries \verb|multcomp|, \verb|tram| and \verb|mlt|, real data can be easily analyzed with this method.

\footnotesize
\bibliographystyle{plain}
\bibliography{RobustDunnett2018}
Acknowledgment: We thank Prof. Dr. Torsten Hothorn (University of Zurich) for the specific instruction for the packages mlt and tram to use for multiple contrast tests.

\end{document}